\documentclass{article}
\usepackage[T1]{fontenc}
\usepackage{iclr2027_conference,times}

\usepackage{amsmath,amsfonts,bm}

\def\eqref#1{equation~\ref{#1}}

\def\1{\bm{1}}

\DeclareMathAlphabet{\mathsfit}{\encodingdefault}{\sfdefault}{m}{sl}
\SetMathAlphabet{\mathsfit}{bold}{\encodingdefault}{\sfdefault}{bx}{n}

\usepackage[hidelinks]{hyperref}
\usepackage{url}
\usepackage{booktabs}
\usepackage{array}
\usepackage{tikz}
\usepackage{bbm}
\usepackage{multirow}
\usepackage{pifont}
\usetikzlibrary{arrows.meta,positioning}

\title{SWE-Residual: Evaluating Starting-State\\Robustness of Coding Agents}
\hypersetup{pdfkeywords={Large Language Models; Coding Agents; Software Engineering; Agent Evaluation; Starting-State Robustness}}

\author{%
  Jinze Li$^{1}$, Weijie Liu$^1$, Jinfeng Xu$^5$, Shuo Yang$^1$, Xin Yang$^3$, Yang Zhang$^4$, Yijun Wang$^2$\\
   \textbf{Zifei Shan$^2$\thanks{Corresponding authors.} \, , \, Edith C.H. Ngai$^{1}$\footnotemark[\value{footnote}]} \\
  $^1$The University of Hong Kong \,
  $^2$Tencent \,  $^3$University of Tokyo \,  $^4$University of North Texas \\
  $^5$The University of British Columbia \\
  \texttt{\{lijinze-hku, liuwj0817, shuo.yang\}@connect.hku.hk} \\
  \texttt{\{zifeishan, ethanyjwang\}@tencent.com} \\
  \texttt{jinfeng.xu@ubc.ca}, \, \texttt{xin.yang@weblab.t.u-tokyo.ac.jp} \\
  \texttt{yang.zhang@unt.edu} , \,
  \texttt{chngai@eee.hku.hk} \\
}

\definecolor{ResultBlue}{RGB}{25,75,150}

\usepackage{etoolbox}
\AtBeginEnvironment{table}{\setlength{\belowcaptionskip}{6pt}}

\iclrfinalcopy
\begin{document}
\maketitle
\pagestyle{plain}  

\begin{abstract}
Software engineering (SWE) benchmarks are the primary way to measure the capabilities of coding agents: an agent is placed in a code repository or machine environment, given a request, and judged by hidden tests. These benchmarks almost always start the agent from a clean state that contains no earlier work on the task. In practice, however, agents often continue work that a developer or another agent has already done and reported as complete, with no indication of what remains incorrect or missing. We introduce SWE-Residual, a benchmark for \emph{Starting-State Robustness}: the ability to maintain performance on a task when the starting state already contains earlier work on that task. SWE-Residual contains 198 Residual States drawn from 111 tasks in DeepSWE and Terminal-Bench, where the earlier work modifies a code repository or a whole machine environment. Each Residual State is the state a predecessor agent left behind after submitting its work as complete, although the hidden tests still fail. The evaluated agent receives only the original request and this state, without the predecessor's trajectory or any sign that the attempt failed; because the request and hidden tests are unchanged, the starting state is the only difference from the source benchmark. Across five frontier agents, success drops by 20.6--53.5 percentage points on DeepSWE and 20.6--34.9 points on Terminal-Bench relative to a Clean Start, and agent rankings on DeepSWE reverse. Failed continuations show a pattern we call \emph{acceptance bias}: agents treat the inherited work as the definition of what remains to be done instead of checking it against the request. On DeepSWE, in 58.9--88.2\% of failed runs, they submit with exactly the same tests failing, although they pass most of these tests from a Clean Start. SWE-Residual thus exposes a capability gap that clean-start evaluation does not capture. Codes are available at https://github.com/userljz/SWE-Residual.
\end{abstract}

\section{Introduction}
\label{sec:introduction}

Coding agents are language models that read files, run commands, and edit
code, and they are now used for software tasks that take hours of work \citep{yang2024sweagent,deng2025swebenchpro,huang2026deepswe}.
As these agents have become more capable, a growing number of benchmarks
have been built to measure their capabilities, from resolving issues in code
repositories to operating a whole machine through the command line
\citep{jimenez2024swebench,huang2026deepswe,merrill2026terminalbench}.
These benchmarks differ in their tasks, but they evaluate an agent in the
same way. The agent is placed in a starting state, such as a code repository or a machine, and given a
request; it works until it declares the task done, and it succeeds if a set
of hidden tests then passes.  Scores obtained in this way are now the main
evidence of how capable a coding agent is.

However, these benchmarks typically start the agent from a clean initial state, one that contains no earlier work on the task the agent is given. We call this a \emph{Clean Start}. Yet in practice, agents often start from states that already contain earlier work on the same task. Development is routinely handed over between developers \citep{khatoonabadi2023wasted}, and increasingly between agents: a user may return in a new session to work that an earlier agent declared done, without the earlier session's context, and in multi-agent systems one agent builds on output that another has submitted as final. In each case, the agent inherits the artifacts of earlier work, such as modified code, newly created files, or installed software. Agents frequently declare success on tasks they have not completed~\citep{advani2026falsesuccess}, so the work handed to the next agent or session often looks finished but is not. We study handoffs in which this work was submitted as complete but does not satisfy the request, and the agent receives no indication of what is incomplete or incorrect: it must work out from the state alone what has been done and what is still missing.

We call the ability to complete a task from a state that already contains such earlier work \emph{Starting-State Robustness}. Most benchmarks cannot measure this ability, because a Clean Start never presents the agent with earlier work, so two agents with the same benchmark score may still differ widely in whether they can finish what someone else has started, and, as we show, their ranking can even reverse. This ability cannot be taken for granted. On one DeepSWE task, for example, an agent inspects the inherited changes, runs the package tests, and submits without changing a single file, although the same agent passes the 11 failing hidden tests from a Clean Start.

\begin{figure}[t]
\centering
\includegraphics[width=\linewidth]{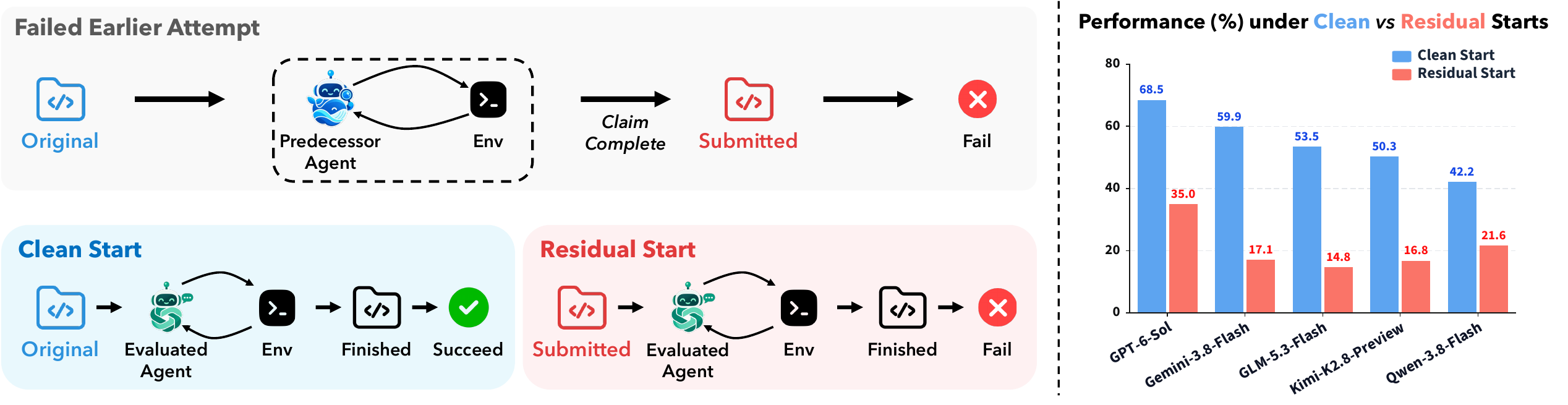}
\caption{\textbf{Left:} Standard benchmarks evaluate an agent from the original repository (\emph{Clean Start}). SWE-Residual evaluates the same agent on the same task from the repository left by a predecessor agent that submitted its work as complete but failed the hidden tests (\emph{Residual Start}), with the same request and hidden tests and no information about the earlier attempt. \textbf{Right:} Success rates of the five evaluated agents under both settings, averaged over DeepSWE and Terminal-Bench.}
\label{fig:protocol}
\end{figure}

To address this gap, we introduce \textbf{SWE-Residual}, a benchmark for
Starting-State Robustness, as shown in Figure~\ref{fig:protocol}.
A \emph{Residual State} is the state a predecessor agent leaves behind after
submitting its work on a task from an existing benchmark as complete, although
the hidden tests still fail; starting an agent from it is a \emph{Residual
Start}. Such a state looks finished: the work is coherent, often passes the
predecessor's own checks, and on DeepSWE the median Residual State already
passes 98.4\% of the hidden tests; yet parts of the request remain missing or
incorrectly implemented, and the evaluated agent must determine which parts
of the apparent solution to trust and what work remains. Each instance of
SWE-Residual pairs a Residual State with the original request and hidden
tests. Because these are unchanged, the same agent can attempt each task
from a Clean Start and from a Residual Start. We report success from
Residual Starts and the change from Clean Start to Residual Start on the
same tasks.

We build SWE-Residual on DeepSWE, where the inherited work modifies a code repository, and Terminal-Bench, where it modifies a machine's files, installed software, and configuration, using Residual States left by predecessor agents from three model families (DeepSeek, Qwen, and GLM). Across five frontier agents, success drops by 20.6--53.5 percentage points on DeepSWE and 20.6--34.9 points on Terminal-Bench relative to Clean Start, and agent rankings on DeepSWE reverse, so strong clean-start performance does not guarantee Starting-State Robustness. We observe a recurring pattern we call \emph{acceptance bias}: the evaluated agent treats the inherited work as the definition of what remains to be done, instead of checking it against the original request. It inspects the inherited changes but keeps most of them and rarely changes the files the predecessor left out. In 58.9--88.2\% of failed continuations on DeepSWE, exactly the same tests fail at the end as at the start, even though the same agent passes most of these tests from a Clean Start.

In summary, our contributions are as follows.
\begin{itemize} 
    \item We identify Starting-State Robustness, the ability to maintain performance on a task when the starting state already contains earlier work on that task, as an evaluation dimension that complements conventional clean-start evaluation. 
    \item We release SWE-Residual, comprising 198 Residual States produced by predecessor agents from three model families (DeepSeek, Qwen, and GLM) on tasks from DeepSWE and Terminal-Bench. We also provide Residualization, a pipeline for constructing and validating these states that enables the benchmark to be extended to additional predecessor agents and source benchmarks.
    \item Across five frontier agents, success drops substantially from a Clean Start to Residual States, and failed continuations show a pattern we call \emph{acceptance bias}: agents treat the inherited work as the definition of what remains to be done instead of checking it against the original request.
\end{itemize}
\section{Related Work}
\label{sec:related-work}
\paragraph{Coding-agent benchmarks.}
Rapid progress in LLMs has spurred substantial research on coding agents for automated software engineering. Such agents are commonly evaluated by placing them in a predefined software environment, providing a task specification, and evaluating whether their resulting changes satisfy execution-based tests. The SWE-bench series~\citep{jimenez2024swebench, deng2025swebenchpro} established this paradigm for resolving real-world repository-level issues, and subsequent benchmarks have extended it to more programming languages~\citep{multiswe2025, rashid2025swepolybench, omnigirl2025}, multimodal inputs~\citep{yang2025swemultimodal}, and longer-horizon tasks~\citep{huang2026deepswe, zhou2026featurebench, miserendino2025swelancer, le2026sweevo}. Beyond code repositories, Terminal-Bench~\citep{merrill2026terminalbench} extends agent evaluation to whole machine environments in which agents must manipulate files, dependencies, configurations, and command-line tools. A related line gives an agent a sequence of requests on one repository and studies how the code accumulated for earlier requests affects later ones~\citep{chen2026sweci,lam2026swechain,shastry2026swesteps,orlanski2026slopcodebench}. Despite differing in task scope and environment, these benchmarks share the assumption that each task begins from a clean starting state containing no prior work on the task itself. This leaves performance on states modified by an earlier attempt at the same task unmeasured.

\paragraph{Benchmarks beyond clean-start evaluation.}
Recent research has begun to relax the clean-start assumption by evaluating coding agents in code or machine environments already modified by prior attempts.  Handoff Debt~\citep{kc2026handoffdebt} and Handoff Tax~\citep{ganz2026handofftax} study takeover from intermediate agent checkpoints, examining how different forms of predecessor context affect the handoff cost of resuming a task. OmniCode~\citep{omnicode} and FeedbackEval~\citep{dai2025feedbackeval} assess whether agents can comprehend and repair code given explicit compiler, test, or LLM feedback. Recovery-Bench~\citep{tan2025recoverybench} replays failed Terminal-Bench runs of a weaker agent and evaluates whether another agent can recover from the resulting environment, finding that more context from the failed attempt can hurt. FixedBench~\citep{gloaguen2026fixedbench} studies the reverse situation, in which the reported issue is already resolved, and finds that agents often make unnecessary changes. Together, these works show that prior attempts can substantially affect subsequent agent behavior and completion cost. Table~\ref{tab:related} summarizes how their settings differ from SWE-Residual, which studies a handoff after a predecessor has submitted its work as complete, although the original task remains unsolved, and pairs each Residual State with a Clean Start of the same agent on the same task.

\paragraph{Relation to Recovery-Bench.}
Recovery-Bench asks whether an agent can recover from a failed attempt: its states come from failed Terminal-Bench runs, whether or not the predecessor submitted its work, and the task is framed as recovery from that attempt. SWE-Residual instead keeps only final states that were submitted as complete and still fail the original verifier; the submission is a selection criterion, and the evaluated agent receives only the request and the inherited state, without the submission or the verdict. The question therefore shifts from recovering after a known failure to finding, in work that appears done, what the original request still requires.

\paragraph{Why agents accept flawed work.}
Our analysis of acceptance bias relates to known limitations of language models. Without external feedback, models often fail to correct their own reasoning~\citep{huang2024selfcorrect,kamoi2024selfcorrectionsurvey}, and they locate errors less reliably than they fix them once the location is given~\citep{tyen2024finderrors}. They also struggle to notice what is missing from an input~\citep{fu2025absencebench}, and they tend to build on earlier mistakes instead of revisiting them~\citep{zhang2024snowball,laban2026lostmultiturn}, even declaring success on tasks they have not completed~\citep{advani2026falsesuccess}. Human developers who reuse an existing artifact show a similar anchoring effect~\citep{parsons2004anchoring}. SWE-Residual measures what these tendencies cost when one agent continues another agent's work. 

\begin{table}[t]
\centering
\footnotesize
\setlength{\tabcolsep}{2.5pt}
\caption{A comparison of SWE-Residual to existing coding
agent benchmarks. \emph{Prior work}: the starting state contains residual artifacts from another agent's earlier attempt. \emph{Submitted as done}: the predecessor submitted the prior work as complete, rather than being interrupted or selected without regard to submission. \emph{Failed}: the prior work does not fully satisfy the task specification. \emph{Blind continuation}: the evaluated agent receives only the resulting state and task request, with no explicit feedback, trajectory, or results of the prior attempt. \ding{51}~yes,  \ding{55}~no, --~not applicable.}
\label{tab:related}
\begin{tabular}{@{}lccccc@{}}
\toprule
Benchmark & Environment & Prior work & \begin{tabular}[c]{@{}c@{}}Submitted\\as done\end{tabular} & Failed & \begin{tabular}[c]{@{}c@{}}Blind\\continuation\end{tabular} \\
\midrule
SWE-bench~\citep{jimenez2024swebench} & Repository & \ding{55} & -- & -- & --  \\
Terminal-Bench~\citep{merrill2026terminalbench} & Machine & \ding{55}& -- & -- & -- \\
\midrule
Handoff Debt~\citep{kc2026handoffdebt} & Repository & \ding{51}& \ding{55} & \ding{55} & \ding{55}  \\
Handoff Tax~\citep{ganz2026handofftax} & Repository & \ding{51}& \ding{55} & \ding{55} & \ding{55}  \\
FeedbackEval~\citep{dai2025feedbackeval} & Repository & \ding{51}& \ding{55} & \ding{51} & \ding{55} \\
OmniCode~\citep{omnicode} & Repository & \ding{51}& \ding{55} & \ding{51} & \ding{55}  \\
Recovery-Bench~\citep{tan2025recoverybench} & Machine & \ding{51}& \ding{55} & \ding{51} & \ding{55} \\
\midrule
\textbf{SWE-Residual (ours)} & Repository \& Machine & \ding{51}& \ding{51} & \ding{51} & \ding{51} \\
\bottomrule
\end{tabular}
\end{table}

\section{SWE-Residual}
\label{sec:method}

This section formalizes the SWE-Residual evaluation setting and describes how we construct and validate its starting states.

\subsection{Task Formulation}
\label{sec:formulation}

\paragraph{Clean Start.}
An evaluation of task \(T_i\) consists of a request \(q_i\), a starting state \(S_i\), and a verifier \(V_i\). The starting state describes the environment that the
agent can observe and modify when it begins: in DeepSWE, this is the code
repository; in Terminal-Bench, it is the entire machine environment,
including files, installed software, configurations, and running processes.
The verifier $V_i$ evaluates the final state using a set of hidden tests that are never exposed to the agent.

For an agent \(A\), let \(F_A(q_i,S_i)\) denote the final state produced by agent \(A\) in a run when the agent either declares the task complete or exhausts the evaluation budget. 
We define the outcome of the run as
\begin{equation}
    y_{A,i}(S_i)
    =
    V_i\left(q_i,F_A(q_i,S_i)\right)
    \in \{0,1\},
\label{eq:success}
\end{equation}
where \(1\) indicates success and \(0\) otherwise. We denote the clean starting state of task $i$ by $S_{0,i}$,
which contains no work from a prior attempt on \(q_i\). A run under the standard clean-start setting therefore has outcome \(y_{A,i}(S_{0,i})\).

\paragraph{Residual States.}
A \emph{Residual State} is the final state left by a prior attempt at the
task that was submitted as complete but failed. Let $\tilde{A}$ denote the predecessor agent that attempts task $T_i$ from
the clean state $S_{0,i}$. We define a Residual State as
\begin{equation}
    S_{r,i}=F_{\tilde{A}}(q_i,S_{0,i}),
    \quad \text{where $\tilde{A}$ declares the task complete and} \
    y_{\tilde{A},i}(S_{0,i})=0.
\label{eq:residual-state}
\end{equation}
That is, $S_{r,i}$ is the state left by the predecessor agent $\tilde{A}$ that attempted the
same request $q_i$ and submitted its work as complete, but did not pass the hidden tests.

\paragraph{Starting-State Robustness.}
We use \emph{Starting-State Robustness} to refer to an agent's ability to
maintain its performance when the same request $q_i$ is attempted from a Residual
State $S_{r,i}$ rather than the clean state $S_{0,i}$. The verifier is held fixed, so the starting state is the only
change.

\subsection{Benchmark Construction}
\label{sec:construction}

We call our process for constructing validated Residual States from attempts by predecessor agents \emph{Residualization}. Residualization consists of four stages. We apply this process to DeepSWE \citep{huang2026deepswe} and Terminal-Bench \citep{merrill2026terminalbench}, while keeping each task's request and verifier unchanged. 

\paragraph{Stage I: Trajectory generation.}
We use three predecessor agents based on DeepSeek-V4-Flash-0731, Qwen-3.8-27B, and GLM-5.3: recent open-weight models of different sizes that are inexpensive to run at scale and capable enough that their work resembles what agents hand over in practice. For each task $T_i$, a predecessor agent $\tilde{A}$ starts from the clean state $S_{0,i}$ and attempts the request $q_i$ under a generous time limit of eight hours per run, producing a final state $F_{\tilde{A}}(q_i,S_{0,i})$ with outcome $y_{\tilde{A},i}(S_{0,i})$. For each run, we record the full trajectory, termination reason, whether the predecessor agent declared completion, and the verifier outcome, together with the information needed to reconstruct its final state: a binary-safe patch of all repository changes on DeepSWE, and the ordered sequence of terminal interactions on Terminal-Bench.

\paragraph{Stage II: Automatic filtering.}
Not every generated run yields a suitable candidate Residual State: some make no persistent changes, some terminate because of infrastructure failures, and others exhaust the time limit before the agent declares the task complete. We therefore apply rule-based filtering before detailed human review. For each generated run, we retain only those that satisfy three criteria: \textbf{(a) Submitted but failing:} the predecessor agent declares the task complete, but $y_{\tilde{A},i}(S_{0,i})=0$. \textbf{(b) Changed:} the final state $F_{\tilde{A}}(q_i,S_{0,i})$ differs from $S_{0,i}$ after removing incidental execution artifacts such as build outputs, caches, and logs. \textbf{(c) No infrastructure failure:} the run does not terminate because of failures unrelated to the task, such as model-request errors, container crashes, or unavailable network resources. Duplicate final states produced for the same task are retained only once.

\paragraph{Stage III: Human review.}
Agent trajectories can span hundreds of steps, making direct step-by-step inspection impractical. Two human reviewers therefore independently assess each candidate using the request $q_i$, the changes from the clean state $S_{0,i}$ to the candidate state $F_{\tilde{A}}(q_i,S_{0,i})$, and a GLM-5.3 summary of the predecessor trajectory that describes the intentions behind the agent's actions and its overall solution strategy. A candidate is retained only if it satisfies two criteria: \textbf{(a) Task-directed:} the summarized strategy reflects a coherent attempt to satisfy the requirements in $q_i$, rather than unrelated or incidental changes. \textbf{(b) Task--verifier consistent:} the request and hidden tests specify consistent requirements for task completion.

\paragraph{Stage IV: State reconstruction.}
The retained candidates are saved artifacts rather than reusable environments, so we reconstruct each one from the clean state $S_{0,i}$ and store it in a reproducible form. On DeepSWE, we apply the saved patch at the task's base commit without creating a commit, so the evaluated agent observes the modified working tree without the commit history that produced it. On Terminal-Bench, we replay the recorded interactions in a fresh task environment and checkpoint the resulting machine state, including live processes. We validate that each reconstructed state remains usable and that the original verifier still returns $V_i(q_i,S_{r,i})=0$ (Appendix~\ref{app:transfer}). During evaluation, each rollout starts from an independent instance of $S_{r,i}$ and receives only the original request $q_i$; the prior trajectory, predecessor outputs, and failure information are not explicitly supplied. We refer to this protocol as \emph{blind continuation}.

\section{Benchmark Characteristics}
\label{sec:characteristics}

SWE-Residual contains 198 validated Residual States derived from 111 tasks across DeepSWE and Terminal-Bench. Each state comes from a prior attempt submitted as complete, passes our inclusion review, and is verified to fail the original task verifier after reconstruction. Table~\ref{tab:dataset} summarizes the benchmark's coverage, predecessor composition, prior-attempt effort, and initial test outcomes.

\begin{table}[t]
\centering
\caption{Statistics of SWE-Residual. Task counts refer to unique source tasks. Prior-attempt steps and execution time are medians over retained states; steps count recorded agent turns. Test pass rates include all tests used for task acceptance and are summarized with equal weight per state.}
\label{tab:dataset}
\begin{tabular}{@{}lcc@{}}
\toprule
 & DeepSWE & Terminal-Bench \\
\midrule
Source benchmark tasks & 113 & 89 \\
Source tasks covered & 90 & 21 \\
Residual States & 165 & 33 \\
\quad from DeepSeek & 62 & 17 \\
\quad from Qwen & 72 & 11 \\
\quad from GLM & 31 & 5 \\
Tasks with one / two / three states
    & 31 / 43 / 16 & 11 / 8 / 2 \\
\midrule
Prior-attempt length (steps) & 134 & 24 \\
Predecessor execution time (min) & 47.7 & 15.5 \\
\midrule
Test pass rate (mean / median, \%)
    & 91.6 / 98.4 & 47.6 / 60.0 \\
States with 1--3 tests still failing
    & 88 (53.3\%) & 33 (100\%) \\
\bottomrule
\end{tabular}
\end{table}

SWE-Residual covers 90 of DeepSWE's 113 tasks, spanning 74 repositories and five programming languages, and 21 of Terminal-Bench's 89 tasks, spanning nine task categories. The retained states reflect substantial prior attempts: the median predecessor run takes 134 agent steps on DeepSWE and 24 on Terminal-Bench, and DeepSWE states change a median of nine files. On Terminal-Bench, the inherited work extends beyond files: retained states include installed software, a compiled application, and running QEMU instances that the evaluated agent must work around or build on. Many states are close to passing: on DeepSWE, the median state passes 98.4\% of the verifier's tests and 53.3\% of states have only one to three tests still failing, yet every retained state fails overall. Appendix~\ref{app:stats} gives further statistics.
\section{Experiments}
\label{sec:experiments}

We first measure how success changes when the starting state moves from the clean state $S_{0,i}$ to a Residual State $S_{r,i}$, across five frontier agents and across predecessor agents. We then ask whether Clean Start successes carry over to Residual Starts, and finally analyze how agents behave when working from Residual States.

\subsection{Experimental Setup}
\label{sec:setup}

\paragraph{Agents and harnesses.}
We evaluate five frontier models as coding agents: Gemini-3.8-Flash, Qwen-3.8-Flash, GLM-5.3-Flash, GPT-6-Sol, and Kimi-K2.8-Preview. We use DeepSWE v1.1 and Terminal-Bench~2.1 as the source benchmarks. Following their evaluation setups, we use mini-swe-agent~\citep{yang2024sweagent} as the agent harness for DeepSWE and Terminus-2~\citep{merrill2026terminalbench} for Terminal-Bench, with all experiments orchestrated through Harbor~\citep{harbor2026}. Each model is evaluated with the highest available reasoning effort. Per-run time limits and other execution budgets follow the default settings of the corresponding source benchmark. Additional details, including sampling parameters, model-specific runtime settings, and infrastructure configurations, are provided in Appendix~\ref{app:eval}.

\paragraph{Metrics.}
Our primary metric is the success rate under Residual Start. For tasks with multiple states, we first average their binary success outcomes within the task, then average across tasks so that each source task receives equal weight. As a matched reference, we report Clean Start success on the same source tasks. The request, verifier, harness, prompt, and evaluation budget are identical between the two conditions; only the starting state changes. We define \emph{Residual Gap} as the Clean Start success rate minus the Residual Start success rate, measured in percentage points. A positive gap indicates lower success under Residual Start. We refer to a run that starts from a Residual State as a \emph{continuation}.

\subsection{Performance on SWE-Residual}
\label{sec:main-results}

We evaluate five frontier agents under blind continuation. Table~\ref{tab:main-results} reports Residual Start success with each source task weighted equally, alongside the matched Clean Start reference and the Residual Gap. Residual Start success ranges from 13.0\% to 33.5\% on DeepSWE and from 12.7\% to 36.5\% on Terminal-Bench. GPT-6-Sol performs best under Residual Start on both tracks, with success rates of 33.5\% and 36.5\%, respectively. Even the highest-scoring agent therefore leaves a large fraction of inherited states unresolved.

\begin{table}[t]
\centering
\caption{Performance on the two SWE-Residual tracks. Residual reports success averaged first over Residual States within each task, then equally across tasks. Ref.\ Clean reports success on the same underlying tasks. Gap is Ref.\ Clean minus Residual, in percentage points, computed before rounding. Bold indicates the highest residual and clean success rates within each track.}
\label{tab:main-results}
\begin{tabular}{@{}lcccc@{}}
\toprule
Track & Agent & Residual (\%) & Ref.\ Clean (\%) & Gap \\
\midrule
\multirow[c]{5}{*}{\begin{tabular}[c]{@{}l@{}}SWE-Residual\\(DeepSWE)\end{tabular}}
    & Gemini-3.8-Flash & 20.0 & \textbf{72.2} & 52.2 \\
    & Qwen-3.8-Flash   & 30.6 & 51.1          & 20.6 \\
    & GLM-5.3-Flash    & 15.4 & 68.9          & 53.5 \\
    & GPT-6-Sol        & \textbf{33.5} & 65.6 & 32.0 \\
    & Kimi-K2.8-Preview & 13.0 & 57.8         & 44.8 \\
\midrule
\multirow[c]{5}{*}{\begin{tabular}[c]{@{}l@{}}SWE-Residual\\(Terminal-Bench)\end{tabular}}
    & Gemini-3.8-Flash & 14.3 & 47.6          & 33.3 \\
    & Qwen-3.8-Flash   & 12.7 & 33.3          & 20.6 \\
    & GLM-5.3-Flash    & 14.3 & 38.1          & 23.8 \\
    & GPT-6-Sol        & \textbf{36.5} & \textbf{71.4} & 34.9 \\
    & Kimi-K2.8-Preview & 20.6 & 42.9         & 22.2 \\
\bottomrule
\end{tabular}
\end{table}

\paragraph{Inheriting earlier work costs every agent 20.6 to 53.5 points.}
Every evaluated agent performs worse under Residual Start than under the matched Clean Start reference. The Residual Gap ranges from 20.6 to 53.5 percentage points on DeepSWE and from 20.6 to 34.9 points on Terminal-Bench. For example, Gemini-3.8-Flash falls from 72.2\% to 20.0\% on DeepSWE, while GPT-6-Sol falls from 71.4\% to 36.5\% on Terminal-Bench.

The decline occurs although the median DeepSWE state already passes 98.4\% of the verifier's tests (Section~\ref{sec:characteristics}), and it appears on both tracks, from inherited repository changes to modified machine environments.

\paragraph{Clean-start rankings reverse on DeepSWE.}
On DeepSWE, Gemini-3.8-Flash and GLM-5.3-Flash rank first and second under Clean Start, but third and fourth under Residual Start, behind GPT-6-Sol and Qwen-3.8-Flash. In particular, GLM-5.3-Flash outperforms Qwen-3.8-Flash by 17.8 percentage points under Clean Start, whereas Qwen-3.8-Flash outperforms GLM-5.3-Flash by 15.2 points under Residual Start. The DeepSWE ranking reversal shows that clean-start and residual-start success can rank agents differently. Absolute residual performance and the drop from Clean Start also capture different aspects of the results: GPT-6-Sol achieves the highest residual success rate on both tracks, while Qwen-3.8-Flash has the smallest Residual Gap.

\paragraph{The drop holds for states from every predecessor agent.}
To examine whether the performance decline persists across predecessor agents, we group all Residual States from DeepSWE and Terminal-Bench by the agent that produced them. Within each group, we compare Residual Start success with the matched Clean Start reference on the same underlying tasks, weighting each task equally. Table~\ref{tab:predecessors} in Appendix~\ref{app:predecessors} shows that all five evaluated agents perform worse under Residual Start in every predecessor group, with Residual Gaps ranging from 19.0 to 51.8 percentage points. The decline therefore occurs across states produced by all three predecessor agents. Since the groups cover different task sets, differences in their absolute scores may also reflect task difficulty.

\subsection{Success Does Not Carry Over from Clean Start}
\label{sec:starting-state-robustness}

\begin{figure}[t]
\centering
\includegraphics[width=\linewidth]{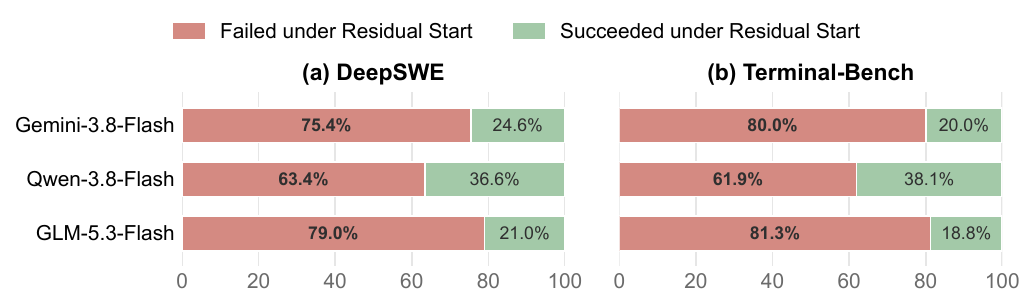}
\caption{Residual Start outcomes on tasks that the same agent successfully completes under Clean Start. Red indicates failure and green indicates success. Outcomes are averaged over Residual States within each task, then equally across tasks. Each agent is evaluated on its own set of observed Clean Start successes.}
\label{fig:starting-state-robustness}
\end{figure}

We next ask whether an agent can still complete a task from a Residual State when it succeeds on the same task from Clean Start. For each agent, we select its observed Clean Start successes and examine its Residual Start outcomes on those tasks. Figure~\ref{fig:starting-state-robustness} shows the resulting failure and success rates.

We run this analysis on three agents: Gemini-3.8-Flash, Qwen-3.8-Flash, and GLM-5.3-Flash. Each has a conditional failure rate above 60\% on both tracks, with rates from 63.4\% to 79.0\% on DeepSWE and from 61.9\% to 81.3\% on Terminal-Bench. For example, Gemini-3.8-Flash fails at a rate of 75.4\% on DeepSWE tasks it completes under Clean Start, while GLM-5.3-Flash fails at a rate of 81.3\% on its corresponding Terminal-Bench tasks.

On tasks each agent completes from a Clean Start, most task-weighted Residual Start outcomes are failures. Success on the original task often fails to carry over once earlier work is inherited. Gemini-3.8-Flash and GLM-5.3-Flash, the strongest of the three under Clean Start, fail most often from inherited states on both tracks.

\subsection{Acceptance Bias}
\label{sec:diagnostics}

From a Clean Start, agents pass most of the tests that a Residual State
fails; from the Residual State itself, most of these tests still fail at the
end. We observe a recurring pattern in their continuations, which we call
\emph{acceptance bias}: once earlier work is present, the agent accepts it
as the definition of what remains to be done, in place of the request. It
continues inside the earlier work and leaves what that work got wrong or
left out as it found it. We examine this on DeepSWE, whose per-test verifier
outcomes, final patches, and reference solutions make each part of this
pattern measurable (Table~\ref{tab:acceptance}).

\begin{table}[t]
\centering
\small
\setlength{\tabcolsep}{4pt}
\caption{Acceptance bias on DeepSWE, with agents sorted by Residual Gap.
\emph{Same failing}: failed continuations that end with exactly the tests
that failed in the Residual State. \emph{Initially failing}: share of those
tests that the agent passes from each start. \emph{Kept}: lines written by
the predecessor agent that remain after failed continuations. \emph{Left-out
files}: runs that change a file changed by the reference solution but not by
the predecessor agent, over states where both runs are normally submitted.
\emph{Steps}: median ratio of Residual to Clean steps. Normally submitted
runs; task-level averages in \% (Gap in points).}
\label{tab:acceptance}
\begin{tabular}{@{}lcccccccc@{}}
\toprule
 & & & \multicolumn{2}{c}{Initially failing} & & \multicolumn{2}{c}{Left-out files} & \\
\cmidrule(lr){4-5}\cmidrule(lr){7-8}
Agent & Gap & Same failing & Clean & Residual & Kept & Clean & Residual & Steps \\
\midrule
GLM-5.3-Flash    & 53.5 & 78.9 & 87.3 & 23.4 & 95.7 & 40.5 & \phantom{0}6.7 & {0.57} \\
Gemini-3.8-Flash & 52.2 & 76.8 & 88.4 & 27.5 & 97.2 & 43.5 & \phantom{0}9.4 & {0.70} \\
Kimi-K2.8-Preview & 44.8 & 88.2 & 77.6 & 14.4 & 98.6 & 45.7 & 8.1 & 0.56 \\
GPT-6-Sol        & 32.0 & 58.9 & 84.1 & 47.7 & 91.1 & 41.2 & 15.0 & 1.02 \\
Qwen-3.8-Flash   & 20.6 & 72.0 & 72.8 & 39.1 & 86.4 & 47.9 & 14.6 & {0.84} \\
\bottomrule
\end{tabular}
\end{table}

\paragraph{Inherited failures survive, although agents can fix them.}
In 58.9\% to 88.2\% of failed continuations, exactly the same verifier tests
fail at the end as at the start. Across failed continuations, agents repair {10.4}\% to {20.1}\% of the
initially failing tests and break almost none of the passing ones. These
tests are not beyond the agents: on the same tasks, the same agent passes
{72.8}\% to {88.4}\% of them from a Clean Start, against 14.4\% to 47.7\% from
the Residual State. The tests become hard only once a failing version of the
work is present.

\paragraph{The earlier work sets the scope of the continuation.}
Agents do not overlook the earlier work: every continuation inspects the
inherited changes, through \texttt{git status}, \texttt{git diff}, or the
modified files. They then work inside it. Discarding the inherited changes
and starting over from the base commit is always possible, and no
continuation does so. Failed continuations keep 86.4\%
to 98.6\% of the lines the predecessor agent wrote, and, except for GPT-6-Sol,
whose ratio is 1.02, continuations take 0.56 to 0.84 times as many steps
as runs from a Clean Start. The clearest
evidence comes from the files that the earlier work left out, that is, files
changed by the reference solution but not by the predecessor agent. On the same
states, agents change these files in {40.5}\% to {47.9}\% of their Clean Start
runs but in only {6.7}\% to {15.0}\% of their Residual Start runs, a difference
of 26.2 to 37.6 points. The files are not unseen: agents open them in {37.6}\%
to {64.0}\% of Residual Start runs, and leave them unchanged.

\paragraph{The more an agent accepts, the more it loses.}
The three agents with the largest gaps, GLM-5.3-Flash, Gemini-3.8-Flash, and
Kimi-K2.8-Preview, show every signature most strongly: they keep the most of the earlier work,
most often leave the same tests failing, change the left-out files least,
and take the fewest steps.
GPT-6-Sol and Qwen-3.8-Flash, whose gaps are 32.0 and
20.6 points, show each signature more weakly. The same holds within agents:
for Gemini-3.8-Flash, Qwen-3.8-Flash, and GLM-5.3-Flash, continuations that
keep more of the earlier work than the agent's median succeed 14.0 to 22.7
points less often than those that keep less.
GPT-6-Sol fails differently. In {11} of 30 sampled failed continuations it names
the defect in the earlier work and tries to repair it, which Gemini-3.8-Flash,
Qwen-3.8-Flash, and GLM-5.3-Flash each do in at most 2 of 30. Among these four
agents, its failed
continuations also repair the most initially failing tests ({20.1}\%, against
10.4\% to 13.7\%) and least often end with the same failures (58.9\%). Its failures
are more often failed repairs than accepted work. Its Residual Gap is
nonetheless 32.0 points: recognizing that the inherited work is wrong recovers
part of the lost performance, and repairing it accounts for the rest.

\paragraph{Case study.}
A run of Gemini-3.8-Flash on \texttt{onedump} shows the pattern in its
plainest form. The Residual State fails 11 verifier tests, all of which the
same agent passes from a Clean Start. From the Residual State, the agent
inspects the inherited changes, runs the package tests, which all pass, and
submits after 43 steps without changing a single file. Its final message
describes the earlier work as its own: ``I have implemented the requested
\texttt{encryption} package.'' The reference solution changes
\texttt{encryption/encryption.go}, a file the predecessor agent never touched;
the evaluated agent leaves it unchanged, while its own Clean Start run
changes it. Qwen-3.8-Flash and GLM-5.3-Flash also fail from this state, each leaving
the same 11 tests failing.

\section{Conclusion}
\label{sec:conclusion}

SWE-Residual measures how well coding agents continue work that a predecessor submitted as complete but that fails the hidden tests, given only the original request and the inherited state. Across five frontier agents, inheriting such work lowers success by 20.6 to 53.5 percentage points on the same tasks, most Clean Start successes do not carry over, and clean-start rankings can reverse. The failures share a pattern we call acceptance bias: agents treat the inherited work as the definition of what remains to be done and leave what it got wrong or left out in place; the more an agent accepts, the more it loses. Clean-start scores therefore overstate how well an agent can take over work in progress, and Starting-State Robustness deserves to be evaluated alongside them. Residualization applies to any benchmark with hidden tests, and SWE-Residual provides a testbed for agents that check inherited work against the request before building on it.

\clearpage
\subsection*{AI Use Statement}
Generative AI tools were used for typo revising, and were not used for method design or experimental analysis. In the benchmark construction, GLM-5.3 generates summaries of predecessor trajectories to support human review (Section~\ref{sec:construction}); all inclusion decisions are made by human reviewers.

\subsection*{Ethics Statement}
Residual States contain unsuccessful implementations produced during
software task attempts. The evaluation protocol uses isolated environments
for executing these research artifacts. Distribution follows the terms of
the source repositories and benchmarks. Upon publication, we will release all
Residual States, the artifacts needed to reconstruct them, the Residualization
pipeline, the evaluation code, and per-run outcomes.

\subsection*{Reproducibility Statement}
Section~\ref{sec:method} specifies the construction and paired evaluation
protocol. Appendix~\ref{app:stats} reports benchmark
statistics, Appendix~\ref{app:transfer} describes state transfer, and
Appendices~\ref{app:eval}--\ref{app:results} give the experimental
configurations and additional analyses. Reconstruction manifests,
filtering records, and per-run outcomes will be publicly released upon
publication.

\bibliography{iclr2027_conference}
\bibliographystyle{iclr2027_conference}

\clearpage
\appendix

\section{Benchmark Statistics}
\label{app:stats}

This appendix expands the summary in Section~\ref{sec:characteristics}; Table~\ref{tab:dataset} lists the main statistics.

\paragraph{Coverage and provenance.}
SWE-Residual covers 90 of DeepSWE's 113 tasks and 21 of Terminal-Bench's 89 tasks. The DeepSWE tasks span 74 repositories and five programming languages: TypeScript, Go, Python, JavaScript, and Rust. The Terminal-Bench tasks span nine official categories, including software engineering, scientific computing, data science, security, and system administration. Both tracks contain states produced by all three predecessor families: DeepSeek, Qwen, and GLM. Of the 111 source tasks, 42 have one retained state, 51 have two, and 18 have three. Tasks with multiple states provide different inherited implementations or environments under the same original request and verifier.

\paragraph{Extent of prior work.}
The retained states reflect substantial prior attempts. The median predecessor run contains 134 agent steps on DeepSWE and 24 on Terminal-Bench, with interquartile ranges of 102--167 and 15--49, respectively. Median agent execution times are 47.7 and 15.5 minutes, including model and tool execution but excluding environment setup and verification. These measurements describe the prior attempts that produced the retained states.

DeepSWE states change a median of nine files, with medians of 1,164 added text lines and 10 deleted text lines. The median patch size is 44.6\,KiB. These changes can include source code, tests, documentation, and configuration. In Terminal-Bench, inherited work can additionally include installed software and running processes; retained examples include a compiled application and running QEMU instances. The median prior attempt contains 45 terminal interactions, including polling and control inputs. The two tracks therefore capture both repository-level changes and work that extends into the machine environment.

\paragraph{Progress in Residual States.}
We characterize each inherited state by the fraction of the original verifier's acceptance tests that it passes before continuation. For DeepSWE, these include both target tests for the requested functionality and regression tests for existing behavior; for Terminal-Bench, they include all task acceptance tests. We compute the pass rate for each state and summarize these rates with equal weight across states. The mean and median are 91.6\% and 98.4\% on DeepSWE, and 47.6\% and 60.0\% on Terminal-Bench, respectively.

On DeepSWE, 88 of 165 states (53.3\%) have one to three tests still failing, while 147 states (89.1\%) preserve all tests that passed in the clean state. On Terminal-Bench, 25 of 33 states (75.8\%) pass at least one test. Twenty-five states have one failing test and eight have two. Of the eight states that pass no tests, seven are evaluated by a single test, which provides no finer-grained measure of partial progress.

These results show that many inherited states already satisfy part of the original verifier, although every retained state still fails overall. The pass rates describe the current state rather than the fraction of work completed by the predecessor agent, since some tests already pass in the clean state. Test suites also differ in granularity and coverage, and even a single failing test may require substantial further work. Our evaluation examines whether agents can identify and complete the work that remains.

\section{Residual State Construction and Transfer}
\label{app:transfer}

This appendix describes how each Residual State is reconstructed from a predecessor run and what the evaluated agent can observe when it starts from that state. Under both Clean Start and Residual Start, the evaluated agent receives the same request, system prompt, harness, and budget; only the starting state differs.

\subsection{DeepSWE}

During Stage~I, we save all changes that the predecessor agent made relative to the task's base commit as a binary-safe patch, covering modified, added, and deleted files, including binary files. The patch retains everything the predecessor left in the repository, including build outputs, caches, and logs; such incidental artifacts are ignored only when Stage~II decides whether a run made substantive changes, and they are not removed from the reconstructed state. To reconstruct the state, we start from the task's original Docker image at the base commit and apply the patch to the working tree without creating a commit. The Git history therefore matches the clean state. The inherited changes appear as uncommitted modifications, visible through commands such as \texttt{git status} and \texttt{git diff}, and any commits the predecessor made during its run are flattened into these modifications, so their commit messages are not transferred. We impose no restriction on Git operations, so an evaluated agent can discard the inherited changes and reimplement the task from the base commit, within the same budget as any other strategy. None of the recorded DeepSWE continuations discarded the inherited work in its entirety.

Files that the predecessor wrote inside the repository, including its own tests, documentation, and notes, remain part of the Residual State, since they are part of the earlier work. The hidden tests used by the verifier are stored outside the repository and are never exposed to the evaluated agent. Tests written or modified by the predecessor remain in the working tree, and the evaluated agent may inspect or change them. Final outcomes are determined by the source benchmark's verifier and acceptance tests, independently of any tests modified by either agent.

\subsection{Terminal-Bench}

In Terminal-Bench, earlier work can change the whole machine, including files, installed packages, configuration, and running processes. During Stage~I, we record the ordered sequence of terminal commands and interactions sent by the predecessor agent. To reconstruct the state, we replay this sequence in a fresh task environment and create a Docker checkpoint using CRIU\footnote{\url{https://criu.org}}, which preserves the container's filesystem and live process state. Each continuation restores an independent instance from this checkpoint. For states with a running QEMU instance, such as those described in Section~\ref{sec:characteristics}, we verify after restoration that the VM process and task-relevant services remain operational.

The evaluated agent receives the original request and the restored machine state. Shell history and other artifacts remain accessible as part of that state; explicit handoff notes and verifier feedback are absent.

\subsection{Validation}

Each reconstructed state is validated before evaluation. A state is usable if it can be instantiated as an independent environment for a rollout: the harness can connect to it, the agent can execute commands and work on the task, and the original verifier can be run on the resulting state. We also check that the original verifier still fails on it, i.e., $V_i(q_i, S_{r,i}) = 0$. A failure must be a valid verdict, meaning that the verifier executed the tests and returned a result; a zero reward caused by an infrastructure or verification error is not treated as a failed attempt, either for a predecessor run or for a reconstructed state. Table~\ref{tab:reconstruction} reports how many candidates were excluded at each step. Every rollout starts from an independent copy of the validated state.

\begin{table}[htbp]
\centering
\small
\caption{Reconstruction and validation of the candidates that passed human review.}
\label{tab:reconstruction}
\begin{tabular}{@{}lccc@{}}
\toprule
 & DeepSWE & Terminal-Bench & Total \\
\midrule
Passed human review & 170 & 37 & 207 \\
\quad Patch application or command replay failed & 2 & 1 & 3 \\
\quad Environment unusable for rollouts & 1 & 2 & 3 \\
\quad No valid verifier result, or no longer failing & 2 & 1 & 3 \\
\midrule
Included in SWE-Residual & 165 & 33 & 198 \\
\bottomrule
\end{tabular}
\end{table}

\subsection{Information Available to the Evaluated Agent}

Table~\ref{tab:transfer} summarizes what the evaluated agent receives under each condition. The two conditions differ only in the starting state and in the earlier work it contains.

\begin{table}[htbp]
\centering
\small
\caption{Information available to the evaluated agent under Clean Start and Residual Start. Both conditions use the same request, system prompt, harness, and budget.}
\label{tab:transfer}
\begin{tabular}{@{}p{0.34\linewidth}p{0.26\linewidth}p{0.32\linewidth}@{}}
\toprule
Item & Clean Start & Residual Start \\
\midrule
Starting state & Original repository or machine & Residual State \\
Request & Original request & Original request \\
System prompt, harness, and budget & Source benchmark defaults & Source benchmark defaults \\
Earlier work inside the state & None & Retained, including code, tests, documentation, and notes written by the predecessor \\
Predecessor trajectory, summaries, or handoff notes & -- & Not explicitly supplied; traces left in the state, such as shell history, may be accessible \\
Indication that an earlier attempt exists or failed & -- & Not provided \\
Stage~III review materials (trajectory summaries) & -- & Not provided \\
Hidden tests and reference solution & Not provided & Not provided \\
\bottomrule
\end{tabular}
\end{table}

\section{Evaluation Configuration and Metrics}
\label{app:eval}

\subsection{Agents and Harnesses}

Table~\ref{tab:config} lists the models used as predecessor agents and as evaluated agents. Evaluated agents use mini-swe-agent on DeepSWE and Terminus-2 on Terminal-Bench, with all runs orchestrated through Harbor, following the evaluation setups of the source benchmarks. Predecessor agents use the same harnesses in Stage~I. Evaluated agents run under the default per-run budget of each source benchmark, identical for Clean Start and Residual Start. Predecessor agents instead run under a generous time limit of eight hours per run, so that as many attempts as possible reach a submission.

We distinguish task failures from infrastructure failures. If an evaluated agent exhausts its budget without solving the task, the run counts as a task failure. If a run is interrupted by an infrastructure problem, such as a transient network, model-service, or container failure, we restart it from an independent instance of the same initial state. If verification fails for infrastructure reasons after the agent has finished, we rerun verification on the saved final state when possible and otherwise repeat the run from its initial state. A verifier error is never recorded as a task failure. The same policy applies to Clean Start and Residual Start.

\begin{table}[htbp]
\centering
\small
\caption{Models used in SWE-Residual. Predecessor agents produce Residual States in Stage~I; evaluated agents are run from Clean Starts and Residual Starts.}
\label{tab:config}
\begin{tabular}{@{}lll@{}}
\toprule
Role & Model & Reasoning and sampling settings \\
\midrule
Predecessor & DeepSeek-V4-Flash-0731 & Highest reasoning effort; default sampling (temperature 1.0) \\
Predecessor & Qwen-3.8-27B & Highest reasoning effort; default sampling (temperature 1.0) \\
Predecessor & GLM-5.3 & Highest reasoning effort; default sampling (temperature 1.0) \\
\midrule
Evaluated & Gemini-3.8-Flash & Highest reasoning effort; default sampling (temperature 1.0) \\
Evaluated & Qwen-3.8-Flash & Highest reasoning effort; default sampling (temperature 1.0) \\
Evaluated & GLM-5.3-Flash & Highest reasoning effort; default sampling (temperature 1.0) \\
Evaluated & GPT-6-Sol & Highest reasoning effort; default sampling (temperature 1.0) \\
Evaluated & Kimi-K2.8-Preview & Highest reasoning effort; default sampling (temperature 1.0) \\
\bottomrule
\end{tabular}
\end{table}

\subsection{Metrics}

Let $\mathcal{T}$ be the source tasks covered by a track and $N = |\mathcal{T}|$. For task $i$, let $\mathcal{R}_i$ be its set of Residual States, with $K_i = |\mathcal{R}_i| \in \{1, 2, 3\}$, and let $y_{A,i}(S) \in \{0, 1\}$ be the verifier outcome of a run of agent $A$ from state $S$ (Section~\ref{sec:formulation}). The Residual Start success rate averages outcomes within each task and then across tasks, so that every source task receives equal weight:
\begin{equation}
\mathrm{Res}(A) = \frac{1}{N} \sum_{i \in \mathcal{T}} \frac{1}{K_i} \sum_{S \in \mathcal{R}_i} y_{A,i}(S).
\end{equation}
The matched Clean Start reference uses the same tasks:
\begin{equation}
\mathrm{Clean}(A) = \frac{1}{N} \sum_{i \in \mathcal{T}} y_{A,i}(S_{0,i}).
\end{equation}
The Residual Gap is $100 \cdot (\mathrm{Clean}(A) - \mathrm{Res}(A))$ percentage points, computed before rounding.

\paragraph{Conditional failure rate.} Section~\ref{sec:starting-state-robustness} restricts the analysis to tasks that the agent completes from a Clean Start, $\mathcal{C}_A = \{ i \in \mathcal{T} : y_{A,i}(S_{0,i}) = 1 \}$, and reports
\begin{equation}
\mathrm{CondFail}(A) = \frac{1}{|\mathcal{C}_A|} \sum_{i \in \mathcal{C}_A} \frac{1}{K_i} \sum_{S \in \mathcal{R}_i} \bigl(1 - y_{A,i}(S)\bigr).
\end{equation}
Because $\mathcal{C}_A$ depends on the agent, each agent is evaluated on its own task set.

\paragraph{Predecessor groups.} Table~\ref{tab:predecessors} pools DeepSWE and Terminal-Bench and weights each source task equally, so the Overall column covers all 111 source tasks. For a predecessor family $g$, the group contains the tasks with at least one Residual State from $g$; within each such task, only the states produced by $g$ are averaged, and the Clean Start reference is computed on the same task set. Groups therefore cover different tasks, and their absolute scores are not directly comparable.

\section{Measuring Acceptance Bias}
\label{app:bias}

The analysis in Section~\ref{sec:diagnostics} is limited to DeepSWE, where per-test verifier outcomes, final patches, and reference solutions are available for every run.

\subsection{Run Selection}

Table~\ref{tab:acceptance} uses normally submitted runs, i.e., runs that end with the agent declaring the task complete and for which the verifier returns a valid result. These include both successful runs and valid failures, and exclude runs that exhaust the budget or end in an infrastructure failure. In the success rates of Table~\ref{tab:main-results}, runs that exhaust the budget count as task failures. Each metric is computed per run, averaged within each task, and then averaged across tasks. Steps is the exception and is reported as a median ratio.

\subsection{Metric Definitions}

For a state $S$, let $\mathcal{F}(S)$ denote the set of verifier tests that fail on $S$. A continuation from a Residual State $S_r$ ends in a final state $S'$ and is a failed continuation if the verifier outcome is 0.

\paragraph{Same failing.} The fraction of failed continuations with $\mathcal{F}(S') = \mathcal{F}(S_r)$, i.e., that end with exactly the tests that failed in the Residual State.

\paragraph{Initially failing.} For each Residual State $S_r$, the fraction of tests in $\mathcal{F}(S_r)$ that pass at the end of the agent's Clean Start run on the same task (Clean) and at the end of its continuation from $S_r$ (Residual). This column covers all normally submitted runs, including successful ones; the repair rate reported in Section~\ref{sec:diagnostics} is the same quantity restricted to failed continuations.

\paragraph{Kept.} For a continuation from $S_r$, let $P$ be the set of text lines that the predecessor added or replaced relative to the base commit; lines the predecessor deleted are not included. Every DeepSWE Residual State has a non-empty $P$. Kept is the fraction of lines in $P$ that remain in the diff of the final state against the base commit, using exact, order-preserving line matching within each file. Table~\ref{tab:acceptance} reports Kept over failed continuations.

\paragraph{Left-out files.} For a Residual State $S_r$, let $L$ be the set of files changed by the reference solution but not by the predecessor. We consider states with non-empty $L$ for which both the agent's Clean Start run and its continuation from $S_r$ are normally submitted. Clean is the fraction of Clean Start runs that change at least one file in $L$; Residual is the fraction of continuations that do so. A continuation opens a file in $L$ if one of its actions exposes the file's contents to the agent. This includes content-reading commands such as \texttt{cat}, \texttt{sed}, \texttt{head}, and \texttt{tail}, \texttt{grep} or \texttt{rg} output that shows file contents, and editor views recorded by the harness. Commands that only list paths, such as \texttt{ls}, \texttt{find}, \texttt{rg -{}-files}, or \texttt{grep -l}, do not count. Neither do \texttt{git status} and \texttt{git diff -{}-stat}; a \texttt{git diff} counts only if its output shows the file's contents.

\paragraph{Steps.} For each continuation, the ratio of its step count to the step count of the agent's Clean Start run on the same task, where a step is one recorded agent turn, as in Table~\ref{tab:dataset}. We average the ratios over the Residual States of each task and report the median over tasks, using normally submitted continuations.

\paragraph{Inspection of inherited changes.} A continuation inspects the inherited changes if it runs \texttt{git status} or \texttt{git diff}, or reads a file that the predecessor modified.

\paragraph{Split by Kept.} To compare continuations within an agent, we compute Kept for every normally submitted continuation, including successful ones, and split them at the agent's median.

\section{Additional Results}
\label{app:results}

\subsection{Performance by Predecessor Agent}
\label{app:predecessors}

Table~\ref{tab:predecessors} reports the per-predecessor results discussed in Section~\ref{sec:main-results}: Residual Start and matched Clean Start success for each evaluated agent, overall and within the states produced by each predecessor agent.

\begin{table}[htbp]
\centering
\small
\caption{Performance by predecessor agent, pooling DeepSWE and Terminal-Bench. All values are success rates (\%). Each group uses its own matched Clean Start reference, with each source task weighted equally. Overall includes all Residual States; the remaining groups include states produced by the indicated predecessor agent. DeepSeek, Qwen, and GLM denote DeepSeek-V4-Flash-0731, Qwen-3.8-27B, and GLM-5.3, respectively.}
\label{tab:predecessors}
\begin{tabular}{@{}ccccccccc@{}}
\toprule
\multirow{3}{*}{Evaluated agent} & \multicolumn{2}{c}{\multirow{2}{*}{Overall}} & \multicolumn{6}{c}{Predecessor agent} \\
\cmidrule(l){4-9}
 & \multicolumn{2}{c}{} & \multicolumn{2}{c}{DeepSeek} & \multicolumn{2}{c}{Qwen} & \multicolumn{2}{c}{GLM} \\
\cmidrule(lr){2-3}\cmidrule(lr){4-5}\cmidrule(lr){6-7}\cmidrule(l){8-9}
 & Clean & Residual & Clean & Residual & Clean & Residual & Clean & Residual \\
\midrule
Gemini-3.8-Flash & 67.6 & 18.9 & 62.0 & 19.0 & 68.7 & 16.9 & 44.4 & 5.6 \\
Qwen-3.8-Flash   & 47.7 & 27.2 & 40.5 & 21.5 & 51.8 & 32.5 & 38.9 & 8.3 \\
GLM-5.3-Flash    & 63.1 & 15.2 & 57.0 & 11.4 & 62.7 & 15.7 & 33.3 & 2.8 \\
GPT-6-Sol       & 66.7 & 34.1 & 63.3 & 32.9 & 62.7 & 34.9 & 47.2 & 19.4 \\
Kimi-K2.8-Preview & 55.0 & 14.4 & 51.9 & 12.7 & 55.4 & \phantom{0}9.6 & 38.9 & 11.1 \\
\bottomrule
\end{tabular}
\end{table}

\section{Limitations}
\label{app:limitations}

Our measurements cover 198 states from 111 tasks, two source benchmarks, and three predecessor families. Extending the collection to human-produced work and other handoff settings would show how the observed gaps vary with the source of inherited work. The acceptance-bias analysis uses DeepSWE, where per-test outcomes and final patches are available.

\end{document}